\documentclass{ws-ijmpb}
\begin{document}

\markboth{T. Hyart, K. N. Alekseev}
{Nondegenerate parametric amplification in  superlattices}

%
\catchline{}{}{}{}{}
%

\title{NONDEGENERATE PARAMETRIC AMPLIFICATION IN SUPERLATTICES AND THE LIMITS OF STRONG AND WEAK DISSIPATION\\}

\author{TIMO HYART}

\address{Department of Physical Sciences,
University of Oulu\\
P.O.Box 3000,
90014 University of Oulu, Finland}

\author{KIRILL N. ALEKSEEV}

\address{Department of Physics, Loughborough University\\
Loughborough LE11 3TU, United Kingdom}

\maketitle

\begin{history}
\received{Day Month Year}
\revised{Day Month Year}
\end{history}

\begin{abstract}

We develop a semiclassical theory of the nondegenerate parametric amplification in a single miniband of superlattice.
We present the formulas describing absorption and gain of signal and idler fields in superlattice and analyze the limiting
cases of strong and weak dissipation. We show how the well-known Manley-Rowe relations arise in the tight-binding
lattice in the weak dissipation limit. Our results can be applied to an amplification of THz signals in semiconductor
superlattices and a control of nonlinear transport of cold atoms in optical lattices.

\end{abstract}


\section{Introduction}

Terahertz radiation ($0.3-10$ THz) has enormous number
of promising applications in different areas of science and
technology, not least because certain important molecular resonances occur at THz frequencies.
One of the main challenges is to construct a coherent miniature source/amplifier of THz radiation that can operate at room temperature.
\par
The classical works of Esaki and Tsu\cite{Esaki} and Ktitorov {\it et al.}\cite{KSS}  stimulated enormous
activity devoted to the non-linear interaction of a high-frequency electric field with
miniband electrons in dc-biased superlattice. Bloch oscillations
of miniband electrons in a such kind of system can, in principle,
result in an amplification of THz radiation (for a review, see Ref.~\refcite{wackerrew}). The main problem in the
realization of these THz amplifiers is a formation of high-field electric domains inside a superlattice under conditions of
negative differential conductivity\cite{Ridley63,Ign87} (NDC). Electric domains are
destructive for gain at THz frequencies.\cite{savvidis}
Several novel ideas have been recently introduced in order to solve the problem (see  Ref.~\refcite{Hyart08} and
papers cited therein).

\par

Another suggested type of THz amplifiers with a superlattice as an active element is based on a coherent
interaction of alternating electric fields with two commensurate frequencies
(Refs.~\refcite{pavlo}--\refcite{Hyart-Cardiff}).
In recent Letters\cite{Hyart07,Hyart06} we clarified the parametric nature of
this effect and found that it can exist without switching the operation point to the NDC part of the time-average voltage-current characteristic of the
superlattice. This theoretical result allows to expect that the undesirable electric instability can be effectively suppressed
in the case of parametric gain.

\par
The parametric amplifiers are characterized by a time-variation of a reactive parameter
which pumps energy into an amplified signal, if the phases and the frequencies of the signal and parameter are suitably
related.\cite{book-paramp}
In a single miniband superlattice the parametric resonance occurs due to oscillations of the effective mass of
miniband electrons\cite{Hyart07}, which is caused by the pump field
\begin{equation}
E_{\rm pump}(t)=E_{\rm dc}+E_0 \cos \omega t.
\end{equation}
In such {\it degenerate} parametric amplifier there is only one resonator mode whose frequency $\omega_1$
is a harmonic or half harmonic of $\omega$. The degenerate amplification in superlattices has been
analyzed in several physically interesting limits, including the quasistatic\cite{Hyart06,Ale06,romanov-jap06}
($\omega\tau\ll 1$, $\omega_1\tau\ll 1$) and ``semiquasistatic'' limits\cite{shorokhov,Hyart-Cardiff}
($\omega\tau<1$, $\omega_1\tau>1$, $\tau$ is the intraband relaxation time).
\par
The degenerate parametric amplification is a phase-sensitive process.
On the other hand, {\it nondegenerate} parametric amplifiers can provide a phase-insensitive amplification of
a given signal.\cite{book-paramp} In this case there are two
resonator modes with incommensurate frequencies $\omega_1$ and $\omega_2$ satisfying
$\omega_1\pm\omega_2=n\omega$ ($n$ is an integer).  There still exist a few publications devoted to the
nondegenerate parametric amplification in superlattices.\cite{Orlov-Rom81_1,Orlov-Rom81_2}
\par
In this paper, we re-examine the problem of nondegenerate parametric amplification in single miniband of
superlattices and analyze the limits of weak ($\tau\rightarrow\infty$) and strong ($\tau\rightarrow 0$)
dissipation.
In the limit of weak dissipation, we show that our formulas for absorption and gain satisfy the well-known
Manley-Rowe relations.\cite{Manley}  We concentrate mainly on the small-signal amplification and demonstrate that
the net absorptions of weak probe fields in superlattice always consist of two distinct terms describing
the incoherent and parametric mechanisms.  We find that for both frequency relations
$\omega_1\pm\omega_2=n\omega$ and odd $n$ the parametric term is zero in unbiased ($E_{\rm dc}=0$) superlattice.
Our results 
provide a mathematical background also for analysis of the nondegenerate parametric amplification in various other
physical systems demonstrating nonlinear transport in a single energy band.

\section{Main formulas describing the nondegenerate parametric amplification in a superlattice}

The stationary current ($t\gg\tau$) in the tight-binding miniband can be calculated for an arbitrary time-dependence of the electric field $E(t)$ as\cite{wackerrew}
\begin{eqnarray}
j(t)&=&2 j_{\rm peak} {\rm{Im}} F(t),  \nonumber \\
F(t)&=&\frac{1}{\tau}\int_{-\infty}^t dt_1
\exp\bigg\{-(t-t_1)/\tau+i\bigg[\int_{t_1}^t dt_2 \frac{e
E(t_2)d}{\hbar} \bigg] \bigg\},
\end{eqnarray}
where $j_{\rm peak}$ is the peak current density in the Esaki-Tsu current-field characteristic
\begin{equation}
j_{\rm peak}=\frac{N e \Delta d}{4 \hbar} \frac{I_1(\Delta/2k_B
T)}{I_0(\Delta/2k_B T)}, \label{peakcur}
\end{equation}
$\Delta$ is the miniband width, $d$ is
the period of the superlattice potential, $N$ is the density of
carriers in the first superlattice miniband, $I_n(x)$ are the modified Bessel functions
and $T$ is the lattice temperature.

In our case the total electric field acting on miniband electrons is
\begin{equation}
E(t)=E_{\rm pump}(t)+E_1 \cos{(\omega_1
t+\phi_1)}+E_2 \cos{(\omega_2 t+\phi_2)}.
\end{equation}
By straightforward calculation we get
\begin{eqnarray*}
F=\sum_{l,k,m, h, c,j} J_l(\beta) J_{l+h} (\beta)
J_k(\beta_1) J_{k+c}(\beta_1) J_m(\beta_2)J_{m+j}(\beta_2)\\
\hspace{2cm}\times \frac{\exp\bigg[i h \omega t+i c(\omega_1
t+\phi_1)+i j(\omega_2 t+\phi_2) \bigg]}{1-i \omega_B \tau-i l
\omega\tau -i k \omega_1 \tau -i m \omega_2 \tau},
\end{eqnarray*}
where $J_l(x)$ are the Bessel functions, $\beta=eE_0d/\hbar \omega$, $\beta_i=eE_id/\hbar \omega_i$ ($i=1,2$) and  $\omega_B=eE_{dc}d/\hbar$ is the Bloch frequency. All summations are from $-\infty$ to $\infty$.

Thus
\begin{eqnarray}
j(t)&=&\sum_{l,k,m, h, c,j} J_l(\beta) J_{l+h}
(\beta) J_k(\beta_1) J_{k+c}(\beta_1) J_m(\beta_2)J_{m+j}(\beta_2)
\nonumber\\&& \times \bigg[ j_{et}(\omega_B \tau+l\omega \tau+ k \omega_1 \tau+m
\omega_2 \tau)  \cos{\big(h \omega t+c(\omega_1 t+\phi_1)+j(\omega_2
t+\phi_2)\big)} \nonumber
\\&&+K(\omega_B \tau+l\omega \tau + k \omega_1 \tau +m \omega_2 \tau) \sin{\big(h \omega t+c(\omega_1 t+\phi_1)+j(\omega_2
t+\phi_2)\big)} \bigg], \nonumber \\ \label{timedepcur}
\end{eqnarray}
where $j_{et}$ is the Esaki-Tsu current-field characteristic
\begin{equation}
\label{ET}
j_{et}(\omega_B \tau)=j_{\rm peak}\frac{2 \omega_B \tau}{1+(\omega_B \tau)^2}
\end{equation}
and $K$ is connected to the $j_{et}$ by the Kramers-Kronig relations\cite{wackerrew} and is given by
\begin{equation}
K(\omega_B \tau)=j_{\rm peak} \frac{2}{1+(\omega_B \tau)^2}.
\end{equation}
The formula (\ref{timedepcur}) describes the time-dependent current response in the case of trichromatic electric field, and both the dc current $j_{dc}=\langle j(t)\rangle_t$ and the absorptions at different frequencies $\mathcal{A}_{\omega}=\langle j(t) \cos\omega t\rangle_t$ and
$\mathcal{A}_{\omega_i}=\langle j(t) \cos(\omega_it+\phi_i)\rangle_t$ ($i=1,2$) can be calculated from this formula. Here $\langle\ldots\rangle_t$ denotes the time-average. The generated/absorbed power densities at different frequencies are then given by $\mathcal{P}_{\omega}=\mathcal{A}_{\omega}E_0$ and $\mathcal{P}_{\omega_i}=\mathcal{A}_{\omega_i}E_i$ ($i=1,2$).
Absorption and emission correspond to $\mathcal{P}>0$ and $\mathcal{P}<0$, respectively.

\subsection{Absorptions for arbitrary amplitudes of the ac fields satisfying $\omega_1+\omega_2=n\omega$}

By assuming that the mutually incommensurate frequencies satisfy the condition of parametric resonance $\omega_1 + \omega_2=n \omega$, we get from formula (\ref{timedepcur}) the following formulas for the absorptions at different frequencies
\begin{eqnarray}
\mathcal{A}_{\omega}&=&\sum_{l, j, k, m} J_l(\beta) \bigg[ J_{l-jn+1} (\beta)+J_{l-jn-1} (\beta)\bigg]
J_k(\beta_1) J_{k+j}(\beta_1) J_m(\beta_2)J_{m+j}(\beta_2) \nonumber\\&& \times
\bigg[ j_{et} (\omega_B\tau+l\omega\tau+ k \omega_1\tau+m \omega_2\tau)
\frac{\cos{\big(j(\phi_1+\phi_2)\big)}}{2}\nonumber\\&&+K(\omega_B \tau+l\omega \tau+ k \omega_1 \tau+m \omega_2\tau)
\frac{\sin{\big(j(\phi_1+\phi_2)\big)}}{2} \bigg],
\label{la1}
\end{eqnarray}
\begin{eqnarray}
\mathcal{A}_{\omega_1}&=&\sum_{l, j, k, m} J_l(\beta) J_{l-jn} (\beta)
J_k(\beta_1) \bigg[ J_{k+j+1}(\beta_1)+ J_{k+j-1}(\beta_1)\bigg] J_m(\beta_2)J_{m+j}(\beta_2)\nonumber\\&& \times
\bigg[ j_{et} (\omega_B\tau+l\omega\tau+ k \omega_1\tau+m \omega_2\tau)
\frac{\cos{\big(j(\phi_1+\phi_2)\big)}}{2} \nonumber\\&&+K(\omega_B\tau+l\omega\tau+ k \omega_1\tau+m \omega_2\tau)
\frac{\sin{\big(j(\phi_1+\phi_2)\big)}}{2} \bigg]
\label{la2}
\end{eqnarray}
and
\begin{eqnarray}
\mathcal{A}_{\omega_2}&=&\sum_{l, j, k, m} J_l(\beta) J_{l-jn} (\beta)
J_k(\beta_1) J_{k+j} (\beta_1) J_m(\beta_2)\bigg[J_{m+j+1}(\beta_2)+J_{m+j-1}(\beta_2)\bigg] \nonumber\\&& \times
\bigg[ j_{et} (\omega_B\tau+l\omega\tau+ k \omega_1\tau+m \omega_2\tau)
\frac{\cos{\big(j(\phi_1+\phi_2)\big)}}{2} \nonumber\\&&+K(\omega_B\tau+l\omega\tau+ k \omega_1\tau+m \omega_2\tau)
\frac{\sin{\big(j(\phi_1+\phi_2)\big)}}{2} \bigg].
\label{la3}
\end{eqnarray}

\subsection{Absorptions for arbitrary amplitudes of the ac fields satisfying $\omega_1-\omega_2=n\omega$}

Similarly one gets in the case of $\omega_1-\omega_2=n\omega$
the following formulas for the absorptions at different frequencies
\begin{eqnarray}
\mathcal{A}_{\omega}&=&\sum_{l, j, k, m} J_l(\beta) \bigg[ J_{l-jn+1} (\beta)+J_{l-jn-1} (\beta)\bigg]
J_k(\beta_1) J_{k+j}(\beta_1) J_m(\beta_2)J_{m-j}(\beta_2) \nonumber\\&& \times
\bigg[ j_{et} (\omega_B\tau+l\omega\tau+ k \omega_1\tau+m \omega_2\tau)
\frac{\cos{\big(j(\phi_1-\phi_2)\big)}}{2}\nonumber\\&&+K(\omega_B \tau+l\omega \tau+ k \omega_1 \tau+m \omega_2\tau)
\frac{\sin{\big(j(\phi_1-\phi_2)\big)}}{2} \bigg],
\label{la4}
\end{eqnarray}
\begin{eqnarray}
\mathcal{A}_{\omega_1}&=&\sum_{l, j, k, m} J_l(\beta) J_{l-jn} (\beta)
J_k(\beta_1) \bigg[ J_{k+j+1}(\beta_1)+ J_{k+j-1}(\beta_1)\bigg] J_m(\beta_2)J_{m-j}(\beta_2)\nonumber\\&& \times
\bigg[ j_{et} (\omega_B\tau+l\omega\tau+ k \omega_1\tau+m \omega_2\tau)
\frac{\cos{\big(j(\phi_1-\phi_2)\big)}}{2} \nonumber\\&&+K(\omega_B\tau+l\omega\tau+ k \omega_1\tau+m \omega_2\tau)
\frac{\sin{\big(j(\phi_1-\phi_2)\big)}}{2} \bigg]
\label{la5}
\end{eqnarray}
and
\begin{eqnarray}
\mathcal{A}_{\omega_2}&=&\sum_{l, j, k, m} J_l(\beta) J_{l-jn} (\beta)
J_k(\beta_1) J_{k+j} (\beta_1) J_m(\beta_2)\bigg[J_{m-j+1}(\beta_2)+J_{m-j-1}(\beta_2)\bigg] \nonumber\\&& \times
\bigg[ j_{et} (\omega_B\tau+l\omega\tau+ k \omega_1\tau+m \omega_2\tau)
\frac{\cos{\big(j(\phi_1-\phi_2)\big)}}{2} \nonumber\\&&+K(\omega_B\tau+l\omega\tau+ k \omega_1\tau+m \omega_2\tau)
\frac{\sin{\big(j(\phi_1-\phi_2)\big)}}{2} \bigg].
\label{la6}
\end{eqnarray}

\subsection{Manley-Rowe relations in the limit of weak dissipation}

In conditions of parametric resonance $\omega_1+\omega_2=n\omega$ in unbiased superlattice, we get in the limit of weak dissipation $\tau \to \infty$
\begin{eqnarray}
\mathcal{A}_{\omega}
&=&\sum_{k,j}J_{-kn}(\beta) \bigg[ J_{-kn-jn+1} (\beta)+J_{-kn-jn-1} (\beta)\bigg]
J_k(\beta_1) J_{k+j}(\beta_1) J_k(\beta_2)J_{k+j}(\beta_2) \nonumber\\&&
\hspace{2cm}\times j_{\rm peak} \sin{\big(j(\phi_1+\phi_2)\big)},
\end{eqnarray}
\begin{eqnarray}
\mathcal{A}_{\omega_1}&=&\sum_{k,j} J_{-kn}(\beta) J_{-kn-jn} (\beta)
J_k(\beta_1) \bigg[ J_{k+j+1}(\beta_1)+ J_{k+j-1}(\beta_1)\bigg] J_k(\beta_2)J_{k+j}(\beta_2)\nonumber\\&&
\hspace{2cm}\times j_{\rm peak}
\sin{\big(j(\phi_1+\phi_2)\big)}
\label{bal1}
\end{eqnarray}
and
\begin{eqnarray}
\mathcal{A}_{\omega_2}&=&\sum_{k,j} J_{-kn}(\beta) J_{-kn-jn} (\beta)
J_k(\beta_1) J_{k+j} (\beta_1) J_k(\beta_2)\bigg[J_{k+j+1}(\beta_2)+J_{k+j-1}(\beta_2)\bigg] \nonumber\\&&
\hspace{2cm}\times j_{\rm peak} \sin{\big(j(\phi_1+\phi_2)\big)}.
\label{bal2}
\end{eqnarray}
By using the Bessel function relation $J_{n-1}(x)+J_{n+1}(x)=2 n J_n(x)/x$, we get
the following relations
\begin{eqnarray}
\mathcal{P}_{\omega}/n\omega &=& -\mathcal{P}_{\omega_1}/\omega_1, \nonumber \\
\mathcal{P}_{\omega}/n\omega &=& -\mathcal{P}_{\omega_2}/\omega_2, \nonumber \\
\mathcal{P}_{\omega_1}/\omega_1 &=& \mathcal{P}_{\omega_2}/\omega_2,
\label{Manley1}
\end{eqnarray}
where the last equation actually follows from the first two relations. These simple connections between the generated/absorbed power at different frequencies are known as the Manley-Rowe relations and are typical for parametric systems consisting of nonlinear inductors and capacitors\cite{Manley}. Indeed in the limit of weak dissipation, electrons in superlattices are gathered together in the quasimomentum space forming an electron bunch which oscillates coherently in quasimomentum space along a certain ballistic trajectory\cite{bunch}.
Here the effective mass of the electrons in the bunch plays the role of the differential
inductance leading to phase-dependent small-signal
absorption and gain for the probe fields if the condition of parametric resonance is satisfied.

Similarly in conditions of parametric resonance $\omega_1-\omega_2=n\omega$, the Manley-Rowe relations in unbiased superlattice are given by equations
\begin{eqnarray}
\mathcal{P}_{\omega}/n\omega &=& -\mathcal{P}_{\omega_1}/\omega_1 \nonumber \\
\mathcal{P}_{\omega}/n\omega &=& \mathcal{P}_{\omega_2}/\omega_2 \nonumber \\
\mathcal{P}_{\omega_1}/\omega_1 &=& -\mathcal{P}_{\omega_2}/\omega_2.
\label{Manley2}
\end{eqnarray}

\section{Absorption of weak signal and idler fields in the presence of strong pump }

If the pump is strong but the probe fields (signal and idler) are weak, modifications of the voltage-current characteristic of superlattice are determined by the action of the pump field alone. In this case, the problem of electric stability has been considered earlier in the theory of degenerate parametric amplification.\cite{Hyart07,Ale06}
\par
Now we want to find the expressions for absorptions of weak probe fields $\beta_i \ll 1$ ($i=1,2$).
By using the Bessel function expansion [$J_0(x)\approx 1$, $J_{\pm1}(x)\approx \pm x/2$], we find from the formulas
(\ref{la2}), (\ref{la3}), (\ref{la5}) and (\ref{la6}) that absorptions at frequencies $\omega_1$ and $\omega_2$
can be written as
\begin{eqnarray}
\mathcal{A}_{\omega_1}&=&\mathcal{A}^{\rm
incoh}_{\omega_1}+\mathcal{A}^{\rm coh}_{\omega_1} \nonumber\\
\mathcal{A}_{\omega_2}&=&\mathcal{A}^{\rm
incoh}_{\omega_2}+\mathcal{A}^{\rm coh}_{\omega_2}. \label{absnondegb}
\end{eqnarray}

\subsection{Incoherent components of absorption}

The phase-independent incoherent absorptions are always given by
\begin{eqnarray}
\mathcal{A}^{\rm incoh}_{\omega_1}&=&\frac{\beta_1}{4}\sum_{l} J_l^2(\beta) \bigg[j_{et}
(\omega_B \tau+l\omega\tau+ \omega_1\tau)-j_{et} (\omega_B\tau+l\omega\tau-
\omega_1\tau)\bigg] \nonumber \\ \mathcal{A}^{\rm
incoh}_{\omega_2}&=&\frac{\beta_2}{4}\sum_{l} J_l^2(\beta) \bigg[j_{et}
(\omega_B \tau+l\omega\tau+ \omega_2\tau)-j_{et} (\omega_B\tau+l\omega\tau-
\omega_2\tau)\bigg]. \nonumber \\ \label{absnondegicoh}
\end{eqnarray}
Without the pump these equations describe a free-electron absorption in unbiased superlattices\cite{Hyart07}. On the other hand in the presence of the pump field, they can in principle also describe the Bloch gain\cite{Hyart08}.

The phase-dependent coherent absorptions result from the parametric resonance in superlattice miniband\cite{Hyart07} and depend on the relation of the frequencies.

\subsection{Coherent absorptions for $\omega_1+\omega_2=n \omega$}

For $\omega_1+\omega_2=n \omega$ the coherent part of the absorptions can be written as
\begin{eqnarray}
\mathcal{A}^{\rm coh}_{\omega_1}&=&\mathcal{B}^{\rm cos}_{\omega_1} \cos(\phi_1+\phi_2)+\mathcal{B}^{\rm sin}_{\omega_1} \sin(\phi_1+\phi_2) \nonumber\\
\mathcal{A}^{\rm coh}_{\omega_2}&=&\mathcal{B}^{\rm cos}_{\omega_2} \cos(\phi_1+\phi_2)+\mathcal{B}^{\rm sin}_{\omega_2} \sin(\phi_1+\phi_2), \label{absnondeg1coh}
\end{eqnarray}
where
\begin{eqnarray}
\mathcal{B}^{\rm cos}_{\omega_1}&=&\frac{\beta_2}{4}\sum_{l}J_l(\beta)
J_{l+n}(\beta) \bigg[j_{et} (\omega_B\tau+l\omega \tau+
\omega_2 \tau) -j_{et}
(\omega_B\tau+l\omega \tau)\bigg] \nonumber\\&&+\frac{\beta_2}{4}\sum_{l}J_l(\beta)
J_{l-n}(\beta)\bigg[j_{et}
(\omega_B\tau+l\omega \tau)-j_{et} (\omega_B \tau+l\omega \tau-
\omega_2 \tau)\bigg],
 \label{absnondeg1aIcoh}
\end{eqnarray}
\begin{eqnarray}
\mathcal{B}^{\rm sin}_{\omega_1}&=&\frac{\beta_2}{4}\sum_{l}J_l(\beta)
J_{l+n}(\beta)\bigg[K
(\omega_B \tau +l\omega \tau)-K (\omega_B \tau+l\omega \tau+
\omega_2 \tau)\bigg]\nonumber \\&&+\frac{\beta_2}{4}\sum_{l}J_l(\beta) J_{l-n}(\beta)
\bigg[K
(\omega_B \tau +l\omega \tau)
-K(\omega_B \tau+l\omega \tau- \omega_2 \tau)\bigg],
 \label{absnondeg1aIIcoh}
\end{eqnarray}
\begin{eqnarray}
\mathcal{B}^{\rm cos}_{\omega_2}&=&\frac{\beta_1}{4}\sum_{l}J_l(\beta)
J_{l+n}(\beta) \bigg[j_{et} (\omega_B\tau+l\omega \tau+
\omega_1 \tau) -j_{et}
(\omega_B\tau+l\omega \tau)\bigg] \nonumber\\&&+\frac{\beta_1}{4}\sum_{l}J_l(\beta)
J_{l-n}(\beta)\bigg[j_{et}
(\omega_B\tau+l\omega \tau)-j_{et} (\omega_B \tau+l\omega \tau-
\omega_1 \tau)\bigg]
 \label{absnondeg1bIcoh}
\end{eqnarray}
and
\begin{eqnarray}
\mathcal{B}^{\rm sin}_{\omega_2}&=&\frac{\beta_1}{4}\sum_{l}J_l(\beta)
J_{l+n}(\beta)\bigg[K
(\omega_B \tau +l\omega \tau)-K (\omega_B \tau+l\omega \tau+
\omega_1 \tau)\bigg]\nonumber \\&&+\frac{\beta_1}{4}\sum_{l}J_l(\beta) J_{l-n}(\beta)
\bigg[K
(\omega_B \tau +l\omega \tau)
-K(\omega_B \tau+l\omega \tau- \omega_1 \tau)\bigg].
 \label{absnondeg1bIIcoh}
\end{eqnarray}

\subsection{Coherent absorptions for $\omega_1-\omega_2=n\omega$}

On the other hand for $\omega_1-\omega_2=n \omega$ the coherent part of the absorptions can be written as
\begin{eqnarray}
\mathcal{A}^{\rm coh}_{\omega_1}&=&\mathcal{B}^{\rm cos}_{\omega_1} \cos(\phi_2-\phi_1)+\mathcal{B}^{\rm sin}_{\omega_1} \sin(\phi_2-\phi_1) \nonumber\\
\mathcal{A}^{\rm coh}_{\omega_2}&=&\mathcal{B}^{\rm cos}_{\omega_2} \cos(\phi_2-\phi_1)+\mathcal{B}^{\rm sin}_{\omega_2} \sin(\phi_2-\phi_1), \label{absnondeg2coh}
\end{eqnarray}
where
\begin{eqnarray}
\mathcal{B}^{\rm cos}_{\omega_1}&=&\frac{\beta_2}{4}\sum_{l}J_l(\beta)
J_{l+n}(\beta) \bigg[j_{et} (\omega_B\tau+l\omega \tau) -j_{et}
(\omega_B\tau+l\omega \tau-\omega_2 \tau)\bigg] \nonumber\\&&+\frac{\beta_2}{4}\sum_{l}J_l(\beta)
J_{l-n}(\beta)\bigg[j_{et}
(\omega_B\tau+l\omega \tau+\omega_2 \tau)-j_{et} (\omega_B \tau+l\omega \tau)\bigg],
 \label{absnondeg2aIcoh}
\end{eqnarray}
\begin{eqnarray}
\mathcal{B}^{\rm sin}_{\omega_1}&=&\frac{\beta_2}{4}\sum_{l}J_l(\beta)
J_{l+n}(\beta)\bigg[K
(\omega_B \tau +l\omega \tau)-K (\omega_B \tau+l\omega \tau-
\omega_2 \tau)\bigg]\nonumber \\&&+\frac{\beta_2}{4}\sum_{l}J_l(\beta) J_{l-n}(\beta)
\bigg[K
(\omega_B \tau +l\omega \tau)
-K(\omega_B \tau+l\omega \tau+ \omega_2 \tau)\bigg],
 \label{absnondeg2aIIcoh}
\end{eqnarray}
\begin{eqnarray}
\mathcal{B}^{\rm cos}_{\omega_2}&=&\frac{\beta_1}{4}\sum_{l}J_l(\beta)
J_{l+n}(\beta) \bigg[j_{et} (\omega_B\tau+l\omega \tau+
\omega_1 \tau) -j_{et}
(\omega_B\tau+l\omega \tau)\bigg] \nonumber\\&&+\frac{\beta_1}{4}\sum_{l}J_l(\beta)
J_{l-n}(\beta)\bigg[j_{et}
(\omega_B\tau+l\omega \tau)-j_{et} (\omega_B \tau+l\omega \tau-
\omega_1 \tau)\bigg]
 \label{absnondeg2bIcoh}
\end{eqnarray}
and
\begin{eqnarray}
\mathcal{B}^{\rm sin}_{\omega_2}&=&\frac{\beta_1}{4}\sum_{l}J_l(\beta)
J_{l+n}(\beta)\bigg[K
(\omega_B \tau +l\omega \tau+\omega_1 \tau)-K (\omega_B \tau+l\omega \tau)\bigg]\nonumber \\&&+\frac{\beta_1}{4}\sum_{l}J_l(\beta) J_{l-n}(\beta)
\bigg[K
(\omega_B \tau +l\omega \tau-\omega_1 \tau)
-K(\omega_B \tau+l\omega \tau)\bigg].
 \label{absnondeg2bIIcoh}
\end{eqnarray}

If the superlattice is unbiased ($\omega_B=0$) and $n$ is odd, we get for both frequency relations $\omega_1 \pm \omega_2=n \omega$ that $\mathcal{A}^{\rm coh}_{\omega_1}=\mathcal{A}^{\rm coh}_{\omega_2}=0$. Thus the nondegenerate parametric amplification can be achieved in unbiased superlattice only if $n$ is even. This result is expected because the oscillations of the effective electron mass in unbiased superlattice take place only at even harmonics\cite{Hyart07}.

\subsection{The limits of weak and strong dissipation}

In the limit of weak dissipation $\tau \to \infty$, we found that the incoherent absorption (\ref{absnondegicoh}) vanishes.
On the other hand, by using the expressions for the coherent absorptions $\mathcal{A}^{\rm coh}_{\omega_1}$ and
$\mathcal{A}^{\rm coh}_{\omega_2}$ (\ref{absnondeg1coh}) we can show that in this limit they satisfy the last equation in the set of Manley-Rowe relations (\ref{Manley1}). In a similar way, we can also show that the
coherent absorptions (\ref{absnondeg2coh}) lead to the the last equation in Manley-Rowe relations (\ref{Manley2}).

\par
In the limit of strong dissipation $\tau \to 0$ (quasistatic limit), the situation is drastically different. Instead of behaving like a nonlinear reactance,
the superlattice in this limit behaves like a nonlinear resistor.
In the quasistatic limit, the asymptotic techniques (Appendix B in Ref.~\refcite{tucker} and Ref.~\refcite{shorokhov}) can be directly applied to find the formulas for the coherent and incoherent parts of absorptions from their general
expressions.\footnote{Same results of course can be derived within the simple quasistatic theory.\cite{{Ale06,romanov-jap06}}}
The incoherent absorptions become
\begin{eqnarray}
\mathcal{A}^{\rm incoh}_{\omega_1}&=&\frac{E_1}{2} \frac{{\rm d}}{{\rm d} E_{\rm dc}}\langle j_{et}\big(E_{\rm
pump}(t)/E_{cr}\big)\rangle_t,\nonumber \\
\mathcal{A}^{\rm incoh}_{\omega_2}&=&\frac{E_2}{2} \frac{{\rm d}}{{\rm d} E_{\rm dc}} \langle j_{et}\big(E_{\rm
pump}(t)/E_{cr}\big)\rangle_t,
\end{eqnarray}
where $E_{cr}=\hbar/ed\tau$ is the Esaki-Tsu critical field corresponding to the peak current and $j_{et}(x)$ is defined in
(\ref{ET}).
We see that the incoherent absorptions are proportional to the dc differential conductivity, and thus in conditions where the domain formation is suppressed they are always positive.\cite{Ale06}

\par
Similarly the coherent absorptions in the quasistatic limit can also be represented as integrals.
In the case of $\omega_1+\omega_2=n\omega$ we get from formulas (\ref{absnondeg1coh}), (\ref{absnondeg1aIcoh}), (\ref{absnondeg1aIIcoh}), (\ref{absnondeg1bIcoh}) and (\ref{absnondeg1bIIcoh}) that
\begin{eqnarray}
\mathcal{A}^{\rm coh}_{\omega_1}&=&\frac{E_2\cos(\phi_1+\phi_2)}{2} \frac{{\rm d}}{{\rm d} E_{\rm dc}}\langle j_{et}\big(E_{\rm
pump}(t)/E_{cr}\big)\cos(n \omega t)\rangle_t , \nonumber \\
\mathcal{A}^{\rm coh}_{\omega_2}&=&\frac{E_1}{E_2}\mathcal{A}^{\rm
coh}_{\omega_1}.
\label{quasistcoh1}
\end{eqnarray}
Whereas in the case of $\omega_1-\omega_2=n\omega$, formulas (\ref{absnondeg2coh}), (\ref{absnondeg2aIcoh}), (\ref{absnondeg2aIIcoh}), (\ref{absnondeg2bIcoh}) and (\ref{absnondeg2bIIcoh}) are transformed to
\begin{eqnarray}
\mathcal{A}^{\rm coh}_{\omega_1}&=&\frac{E_2\cos(\phi_1-\phi_2)}{2} \frac{{\rm d}}{{\rm d} E_{\rm dc}}\langle j_{et}\big(E_{\rm
pump}(t)/E_{cr}\big)\cos(n \omega t)\rangle_t , \nonumber \\
\mathcal{A}^{\rm coh}_{\omega_2}&=&\frac{E_1}{E_2}\mathcal{A}^{\rm
coh}_{\omega_1}.
\label{quasistcoh2}
\end{eqnarray}
We can make two important remarks concerning these equations. First, they are equivalent to the corresponding equations
in the case of degenerate parametric amplification\cite{Ale06}. Thus we can immediately say that the parametric gain can
overcome the incoherent
absorption leading to amplification of the probe fields in conditions where the formation of electric domains is suppressed.
Secondly, we have an interesting relation for the coherent parts of the generated/absorbed power
$\mathcal{P}^{\rm coh}_{\omega_i}=E_i \mathcal{A}^{\rm coh}_{\omega_i}$ ($i=1,2$). Namely
\begin{equation}
\mathcal{P}^{\rm coh}_{\omega_1}=\mathcal{P}^{\rm coh}_{\omega_2}.
\label{strongdis}
\end{equation}

\par
Finally, we would like to outline the differences in amplification/attenuation of the probe fields for the two mixing relations
$\omega_1\pm\omega_2=n\omega$ in a variation of the dissipation strength.
In the case $\omega_1+\omega_2=n\omega$, the {\it simultaneous amplification} of both modes is possible both in the limit of weak (\ref{Manley1}) and strong dissipation (\ref{strongdis}). The main difference in this case is the change in the phase-dependence of absorptions from sine (\ref{bal1}), (\ref{bal2}) to cosine (\ref{quasistcoh1}) with an increase of dissipation.
In the case $\omega_1-\omega_2=n\omega$, the changes are more drastic. Whereas the simultaneous amplification of both modes is possible in the limit of strong dissipation (\ref{strongdis}), it is forbidden in the limit of weak dissipation (\ref{Manley2}).

\section{Conclusion and perspectives}

We contributed to the theory of nondegenerate parametric amplification in superlattices. We showed
that the solution of Boltzmann transport equation for a tight-binding lattice satisfies the
Manley-Rowe relation in the limit of weak dissipation. It demonstrates that we are really dealing with the parametric
processes. Our theory predicts that in semiconductor superlattices the amplification of electric field can potentially
be reached in a wide range of frequencies from microwaves up to high terahertz.

\par
We believe that similar parametric effects can be realized in various other physical
systems. Here we would like to mention only two of them: Nitride semiconductors and optical lattices.
Since the dependence of effective mass on the energy density
in the lower part of the conduction band of dilute nitride alloys\cite{Ignatov-Patane1,Ignatov-Patane2} resembles the corresponding dependence occurring in the miniband of
superlattices, we speculate that our theories of nondegenerate and degenerate parametric
amplification can be also applicable to these bulk semiconductors. On the other hand,
the tight-binding optical lattices filled with cold atoms or Bose-Einstein condensates can demonstrate nonlinear transport
in a single energy band in the both dissipative\cite{Ott} and nondissipative limits\cite{Madison,Sias}.
Therefore the optical lattices can be considered as a playground for the demonstration of ac-driven transport satisfying
the Manley-Rowe relations. A detailed study of the parametric resonance in these interesting systems
goes beyond the scope of the present paper.

\section*{Acknowledgements}

We are grateful to Alexey Shorokhov and Jussi Mattas for collaboration and Feo Kusmartsev for valuable advices.
This work was partially supported by Vilho, Yrj\"{o} and Kalle V\"{a}is\"{a}l\"{a} Foundation, Emil Aaltonen Foundation and AQDJJ Programme of ESF.

\end{document}